\documentclass[aps,prd,showpacs,floatfix,eqsecnum,amsmath,amssymb,nofootinbib,
twocolumn,showpacs]{revtex4-1}
\usepackage{graphicx}
\usepackage{dcolumn}
\usepackage{bm}
\usepackage{hyperref}
\usepackage{color}
\begin{document}

\title{Constraint Propagation of $C^2$-adjusted Formulation\\
--- Another Recipe for Robust ADM Evolution System ---}

\author{Takuya Tsuchiya}
\email{tsuchiya@akane.waseda.jp}

\author{Gen Yoneda}%
\affiliation{%
Department of Mathematical Sciences, Waseda University, Okubo, Shinjuku,
Tokyo, 169-8555, Japan
}%

\author{Hisa-aki Shinkai}
\affiliation{
Faculty of Information Science and Technology, Osaka Institute of
Technology, 1-79-1 Kitayama, Hirakata, Osaka 573-0196, Japan\\
Computational Astrophysics Laboratory,
Institute of Physical \& Chemical Research (RIKEN),
Hirosawa, Wako, Saitama, 351-0198 Japan
}%
\date{\today}

\begin{abstract}
With a purpose of constructing a robust evolution system against numerical
instability for integrating the Einstein equations,
we propose a new formulation by adjusting the ADM evolution
equations with constraints.
We apply an adjusting method proposed by Fiske (2004) which uses
the norm of the constraints, $C^2$.
One of the advantages of this method is that the effective signature of
adjusted terms (Lagrange multipliers) for constraint-damping evolution is
pre-determined.
We demonstrate this fact by showing the eigenvalues of constraint propagation
equations.
We also perform numerical tests of this adjusted evolution system using
polarized Gowdy-wave propagation, which show robust evolutions against the
violation of the constraints than that of the standard ADM formulation.
\end{abstract}

\maketitle

\section{Introduction}
\label{Introduction}

The standard way for integrating the Einstein equations is to split spacetime
into space and time.
The Arnowitt-Deser-Misner (ADM) formulation \cite{ADM62,York78} provides
the fundamental spacetime decompositions.
However, it is known that the set of the ADM evolution equations is not
appropriate for numerical simulations such as the coalescences of the binary
neutron-stars and/or black-holes, which are the main targets of gravitational
wave sources, and which requires quite long-term time integration.

In order to perform an accurate and stable long-term numerical simulation in
strong gravitational field, we need to modify the ADM evolution equations.
This is called as the ``formulation problem in numerical relativity''
\cite{SY00,SY02gr-qc,Shinkai09}.

The origin of the formulation problem is the violation of constraints, which
triggers the blow-up of simulations.
The discretization of equations arises truncation errors inevitably, so that
we have to adjust the evolution system which is robust for error-growing
modes.
Several formulations are suggested and applied; among them,
the Baumgarte-Shapiro-Shibata-Nakamura (BSSN) formulation \cite{SN95,BS98},
the generalized-harmonic (GH) formulation \cite{PretriusCQG05, Garfinkle02}
and the Kidder-Scheel-Teukolsky (KST) formulation \cite{KST01} are applied
widely for the inspiral black-hole binary mergers.
(Many numerical simulations are reported, but we here cite the works
\cite{BCCKM06, CLMZ06} for applications of the BSSN formulation,
\cite{Pretorius05} for the GH formulation, and \cite{SBCKMP09} for the KST
formulation).
There are also many other formulations which are waiting to be tested
\cite{YS01cqg, YS02, BLPZ03, GCHM05, PHK08}.

The current succeeded large-scale numerical simulations are applying such modern reformulations, but also
using the ``constraint-damping" technique, which is obtained by adding the constraint terms to evolution equations.
The additional constraint-damping terms are reported to be the key implementation in
BSSN and GH system (e.g. \cite{BBCKMM08, Pretorius06}).
We \cite{YS01prd, SY02, YS02} systematically investigated how the additional
constraint terms change the original evolution systems, under the name
``adjusted systems''.
As we will review in Sec.\ref{GeneralIdea}, monitoring the stability of the
evolution is equivalent to check the constraint propagation equations
(dynamical equations of constraints).
Therefore, we proposed to analyze the eigenvalues of the constraint propagation
equations, which can predict the violation of constraints before we try actual
simulations.

Based on the same motivation with this ``adjusted system'', Fiske \cite{Fiske04}
proposed an adjustment which uses the norm of constraints, $C^2$, which
we call the ``$C^2$-adjusted formulation''.
He applied this method to the Maxwell equations, and reported that this method
reduces the constraint violations for a certain range of the coefficient.
An advantage of this $C^2$-adjusted formulation is that the effective signature
of the coefficients is pre-determined.
In this article, we apply the $C^2$-adjusted formulation to the ADM evolution
equations, since the ADM formulation is the one of the most basic evolution
systems in general relativity.
We show the eigenvalue analysis of the constraint propagation of this set, and
also demonstrate numerical evolutions.

Before the numerical relativity groups faced the formulation problem, Detweiler
\cite{Detweiler87} suggested another adjustment based on the ADM evolution
equations.
He proposed a particular combination of adjustments which make the norm of
constraints damp down.
The story is quite similar to this work.
However, Detweiler's method is restricted with the maximal slicing condition,
$K=0$, and also the behavior except the flat-space is unknown.
We also show numerical demonstrations of Detweiler's evolution equation for
a comparison.

We compare the violations of the constraints between the standard ADM,
Detweiler's ADM and $C^2$-adjusted ADM formulations.
We use the polarized Gowdy-wave evolution which is one of the comparison test
problems as is known to the Apples-with-Apples testbeds
\cite{Alcubierre_etc04}.
The models precisely fixed up to the gauge conditions, boundary conditions, and
technical parameters,
therefore testbeds are often used for comparison between formulations
\cite{Zumbusch09, BB10, KS08}.

The plan of this article is as follows.
We review the idea of adjusted systems and $C^2$-adjusted formulation in
Sec.\ref{GeneralIdea}.
We also describe a recipe for analyzing the constraint propagation with its
eigenvalue analysis which we call  the constraint amplification factors (CAFs).
In Sec.\ref{ApplicationEinsteinEq}, we apply the $C^2$-adjusted formulation to
the ADM equations and show its CAFs.
We also review Detweiler's formulation in Sec.\ref{GeneralIdea}.
We show our numerical evolutions
in Sec.\ref{NumericalExamples}, and we summarize this article in
Sec.\ref{Summary}. In this article, we only consider the vacuum spacetime, but
the inclusion of matter is straightforward.


\section{The idea of adjusted systems and $C^2$-adjusted systems}
\label{GeneralIdea}

\subsection{The idea of adjusted systems}
\label{GeneralIdea_AdjustedSystem}

We review the general procedure of rewriting the evolution equations which we
call adjusted systems \cite{YS02, SY02, YS01cqg,YS01prd}.
Suppose we have dynamical variables $u^i$ which evolve along with the evolution
equations,
\begin{align}
&\partial_t u^i = f(u^i, \partial_j u^i, \cdots),
\label{eq:generalEvolveEquations}
\end{align}
and suppose also that the system has the (first class) constraint equations,
\begin{align}
&C^a(u^a, \partial_j u^a, \cdots)\approx 0.
\label{eq:generalConstraintEquations}
\end{align}
We propose to study the properties of the evolution equation of  $C^a$ (which
we call the constraint propagation),
\begin{align}
\partial_t C^a &=g(C^a,\partial_i C^a,\cdots),
\label{eq:generalConstraintPropagation}
\end{align}
for predicting the violation behavior of constraints, $C^a$, in time evolution.
The equation \eqref{eq:generalConstraintPropagation} is theoretically weakly
zero, i.e. $\partial_t C^a \approx 0$, since the system is supposed to be the
first class. However, the free numerical evolution with the discretized grids
introduces constraint violation at least the level of truncation error, which
sometimes grows to stop the simulations.
The set of the ADM formulation has such a disastrous feature even in the
Schwarzschild spacetime, as was shown in \cite{SY02}.

Such features of the constraint propagation equations,
\eqref{eq:generalConstraintPropagation}, will be changed when we modify the
original evolution equations.
Suppose we add the constraint terms to the right-hand-side of
\eqref{eq:generalEvolveEquations} as
\begin{align}
 \partial_t u^i = f(u^i, \partial_j u^i, \cdots) +
F(C^a, \partial_j C^a, \cdots),
\label{eq:generalADjustedEvolutionEqs}
\end{align}
where $F(C^a,\cdots)\approx0$ in principle but not exactly zero in numerical
evolutions,  then \eqref{eq:generalConstraintPropagation} will also be modified
as
\begin{align}
\partial_t C^a&=g(C^a,\partial_i C^a,\cdots) + G(C^a, \partial_i C^a,
\cdots).
\label{eq:generalADjustedConstraintPropagationEqs}
\end{align}
Therefore we are able to control $\partial_t C^a$ by an appropriate adjustment
$F(C^a, \partial_j C^a, \cdots)$ in \eqref{eq:generalADjustedEvolutionEqs}.
There exist various combinations of $F(C^a, \partial_j C^a, \cdots)$ in
\eqref{eq:generalADjustedEvolutionEqs}, and all the alternative formulations are
using this technique.
Therefore, our goal is to find out a better way of adjusting the evolution
equations which realizes $\partial_t C^a\leq 0$.

\subsection{The idea of $C^2$-adjusted formulations}
\label{GeneralIdea_C2Adjusted}

Fiske \cite{Fiske04} proposed an adjustment of the evolution equations in the
way of
\begin{align}
\partial_t u^i = f(u^i, \partial_j u^i, \cdots)-\kappa^{i j}
\left(\frac{\delta C^2}{\delta u^j}\right),
\label{eq:adjutedGeneralEvolveEquations}
\end{align}
where $\kappa^{i j}$ is positive-definite constant coefficient, and $C^2$ is the
norm of constraints which is defined as $\displaystyle{C^2\equiv \int C_a C^a
d^3 x}$.
The term $(\delta C^2/\delta u^j)$ is the functional derivative of $C^2$ with
$u^j$.
We call the set of \eqref{eq:adjutedGeneralEvolveEquations} with
\eqref{eq:generalConstraintEquations} as ``$C^2$-adjusted formulation''.
The associated constraint propagation equation becomes
\begin{align}
\partial_t C^2 = h(C^a, \partial_i C^a, \cdots) - \int d^3 x
\left(\frac{\delta C^2}{\delta u^i}\right)\kappa^{i j}
\left(\frac{\delta C^2}{\delta u^j}\right).
\label{eq:adjustedGeneralConstraintPropagation_of_C2}
\end{align}

If we set $\kappa^{i j}$ so as the second term  in the RHS of
\eqref{eq:adjustedGeneralConstraintPropagation_of_C2} becomes dominant than
the first term, then $\partial_t C^2$ becomes negative, which indicates that
constraint violations are expected to decay to zero.
Fiske presented some numerical examples in the Maxwell system, and concluded
that this method actually reduces the constraint violations.
He also reported that the coefficient $\kappa^{i j}$ has a practical upper
limit in order not to crash simulations.

\subsection{The idea of CAFs}
\label{GeneralIdea_CAFs}
There are many efforts of re-formulation of the Einstein equations which make
the evolution equations in an explicit first-order hyperbolic form (e.g.
\cite{BM92, KST01,BLPZ03, GM04}).
This is motivated by the expectations that the symmetric hyperbolic system has
well-posed properties in its Cauchy treatment in many systems and that the
boundary treatment can be improved if we know the characteristic speed of the
system.
The advantage of the standard ADM system \cite{York78} (compared with the
original ADM system \cite{ADM62}) is reported by Frittelli \cite{Frittelli97}
from the point of the hyperbolicity of the constraint propagation equations.
However, the classification of hyperbolicity(weakly,  strongly or symmetric
hyperbolic) only uses the characteristic part of evolution equations and ignore
the rest.
Several numerical experiments \cite{SY00,Hern00} reported that such a
classification is not enough to predict the stability of the evolution system,
especially for highly non-linear system like the Einstein equations.

In order to investigate the stability structure of
\eqref{eq:generalADjustedConstraintPropagationEqs}, the authors \cite{YS01prd}
proposed the constraint amplification factors(CAFs).
The CAFs are the eigenvalues of the coefficient matrix, $M^a{}_b$ (below), which
is the Fourier-transformed components of the constraint propagation equations,
$\partial_t\hat{C}^a$.
That is,
\begin{align}
&\partial_t \hat{C}^a = g(\hat{C}^a) = M^a{}_b \hat{C}^b,
\nonumber\\
&{\rm where}\,\,\, C^a(x,t) = \int \hat{C}(k,t)^a \exp({\rm i} k\cdot x)
 d^3k.
\label{eq:fourierTransformedConstraintEquations}
\end{align}
CAFs include all the contributions of the terms, and enable us to check the
eigenvalues.
If CAFs have negative real-part, the constraints are forced to be diminished.
Therefore, we expect more stable evolution than a system which has CAFs with
positive real-part.
If CAFs have non-zero imaginary-part, the constraints are supposed to propagate
away.
Therefore, we expect more stable evolution than a system which has CAFs with
zero imaginary-part.
The discussion and examples are shown in \cite{SY00, YS01cqg}, where several
adjusted-ADM systems \cite{SY00} and adjusted-BSSN systems \cite{YS02} are
proposed.

\section{Application to the ADM formulation}
\label{ApplicationEinsteinEq}

\subsection{The standard ADM formulation and $C^2$-adjusted ADM formulation}
\label{C2adjustedADMFormulation}
We start by presenting the standard ADM formulation \cite{York78} of the
Einstein equations.
The standard ADM evolution equations are written as
\begin{align}
\partial_t \gamma_{i j}
&=-2\alpha K_{i j} + D_i \beta_j + D_j \beta_i,
\label{eq:gamma_standardADMEvolutionEquations}\\
\partial_t K_{i j}
&= \alpha({}^{(3)}R_{i j} + K K_{i j} - 2K_{i \ell} K^\ell{}_j)
- D_i D_j \alpha\nonumber\\
&\quad
+ K_{\ell i} D_j \beta^\ell + K_{\ell j} D_i \beta^\ell
 + \beta^\ell D_\ell K_{i j},
 \label{eq:extrinsicCurvature_standardADMEvolutionEquations}
\end{align}
where $(\gamma_{i j}, K_{i j})$ are the induced three-metric and the extrinsic
curvature, $(\alpha, \beta^i)$ are the lapse function and the shift vector,
$D_i$ is the covariant derivative associated with $\gamma_{i j}$ and
${}^{(3)}R_{i j}$ is the three Ricci tensor.
The constraint equations are
\begin{align}
\mathcal{H}
& \equiv {}^{(3)}R + K^2 - K_{i j} K^{i j} \approx 0,
\label{eq:hamiltonian_standardADMConstraintEquations}\\
\mathcal{M}_i
& \equiv D_j K^j{}_i - D_i K \approx 0,
\label{eq:momentum_standardADMConstraintEquations}
\end{align}
where ${}^{(3)}R$ is the three-scalar curvature, ${}^{(3)} R = \gamma^{i j}
{}^{(3)}R_{i j}$ and $K$ is the trace-part of the extrinsic curvature, $K =
\gamma^{i j} K_{i j}$.

The constraint propagation equations of the Hamiltonian constraint,
\eqref{eq:hamiltonian_standardADMConstraintEquations},
and the momentum constraints,
\eqref{eq:momentum_standardADMConstraintEquations}, can be written as
\begin{align}
\partial_t \mathcal{H}
& = \beta^i D_i \mathcal{H} - 2 \alpha D_i \mathcal{M}^i
+ 2 \alpha K \mathcal{H}
- 4 (D_i \alpha) \mathcal{M}^i,
\label{eq:hamiltonian_constraintPropagationEquations_of_ADM}\\
\partial_t \mathcal{M}_i
& = - (1/2) \alpha D_i \mathcal{H} + \beta^\ell D_\ell \mathcal{M}_i
- (D_i \alpha) \mathcal{H}\nonumber\\
&\quad
+ (D_i \beta^\ell) \mathcal{M}_\ell + \alpha K\mathcal{M}_i,
\label{eq:momentum_constraintPropagationEquations_of_ADM}
\end{align}
respectively.

Now we apply $C^2$-adjustment to the ADM formulation, which can be written as
\begin{align}
\partial_t \gamma_{i j}
& = \eqref{eq:gamma_standardADMEvolutionEquations} - \kappa_{\gamma i j m n}
\left(\frac{\delta C^2}{\delta \gamma_{m n}}\right),
\label{eq:gamma_c2AdjustedSystemADM}\\
\partial_t K_{i j}
& = \eqref{eq:extrinsicCurvature_standardADMEvolutionEquations} - \kappa_{K i j m n}
\left(\frac{\delta C^2}{\delta K_{m n}}\right),
\label{eq:extrinsicCurvature_c2AdjustedSystemADM}
\end{align}
where $C^2$ is the norm of the constraints, which we set
\begin{align}
C^2 \equiv \int ({\mathcal{H}}^2 + \gamma^{i j}\mathcal{M}_i \mathcal{M}_j)
d^3 x,
\label{eq:norm_of_constraintsADMSystem}
\end{align}
and both coefficients of $\kappa_{\gamma i j m n}, \kappa_{K i j m n}$
are supposed to be positive definite.
We write $(\delta C^2/ \delta \gamma_{m n})$ and $(\delta C^2/ \delta K_{m
n})$ explicitly as \eqref{eq:appendix_ADM_deltaGamma1} and
\eqref{eq:appendix_ADM_deltaK1} in Appendix \ref{Appendix_adjustTermsC2}.

\subsection{Constraint Propagation with $C^2$-adjusted ADM formulation}
\label{constraintPropagation}

In this subsection, we discuss the constraint propagation of the $C^2$-adjusted
ADM formulation, by giving the CAFs on flat background metric.
We show CAFs are negative real numbers or complex numbers with negative
real-part.

The constraint propagation equations,
\eqref{eq:hamiltonian_constraintPropagationEquations_of_ADM} and
\eqref{eq:momentum_constraintPropagationEquations_of_ADM}, are changed due to
$C^2$-adjusted terms.
The full expressions of the constraint propagation equations are shown as
\eqref{eq:app_CPHC2} and \eqref{eq:app_CPMC2} in Appendix
\ref{Appendix_constriantPropagation_C2ADM}.

If we fix the background is flat spacetime, ($\alpha=1, \beta^i=0, \gamma_{i
j}=\delta_{i j}, K_{i j}=0$), then CAFs are easily derived.
For simplicity, we also set $\kappa_{\gamma i j m n} =\kappa_{K i j m n}
=\kappa \delta_{i m}\delta_{j n}$, where $\kappa$ is positive.
The Fourier-transformed equations of the constraint propagation equations are
\begin{align}
&\partial_t
\left(
\begin{array}{c}
\hat{\mathcal{H}}\\
\hat{\mathcal{M}}_i
\end{array}
\right)\nonumber\\
&=\left(
\begin{array}{cc}
-4\kappa|\vec{k}|^4 & -2{\rm i}k_j\\
-(1/2){\rm i}k_i & \kappa( -|\vec{k}|^2\delta_{i j} -3k_i k_j)
\end{array}
\right)
\left(
\begin{array}{c}
\hat{\mathcal{H}}\\
\hat{\mathcal{M}}_j
\end{array}
\right).
\label{eq:fourierTransformedConstraintPropagation_Of_C2AdjustedADM}
\end{align}

The eigenvalues, $\lambda$,  of the coefficient matrix of
\eqref{eq:fourierTransformedConstraintPropagation_Of_C2AdjustedADM} are given
by solving
\begin{align*}
(\lambda+\kappa |k|^2)^2(\lambda^2+A\lambda +B)=0,
\end{align*}
where $A\equiv 4\kappa |k|^2 (|k|^2+1)$ and $B\equiv |k|^2+16\kappa^2 |k|^6$.
Therefore, the four eigenvalues are 
\begin{align}
(-\kappa |k|^2, -\kappa |k|^2, \lambda_+ ,  \lambda_-), 
 \label{CAF_C2ADM}
\end{align}
where 
\begin{align}
 \lambda_\pm = -2\kappa|k|^2 (|k|^2+1)\pm |k|\sqrt{-1+4\kappa^2
 |k|^2(|k|^2-1)^2}.
 \label{lambda_pm}
\end{align}
From the relation of the coefficients with solutions,
\begin{align}
\lambda_++\lambda_- = - A < 0,
\quad{\rm and}\quad
\lambda_+\lambda_-=B > 0,\label{eq:C2CAF}
\end{align}
we find both the real parts of $\lambda_+$ and $\lambda_-$ are negative.
Therefore, we see all four eigenvalues are complex numbers with negative real-part or
negative real numbers.

On the other hand, the CAFs of the standard ADM formulation on flat background 
[$\kappa=0$ in (\ref{CAF_C2ADM})]
are reduced to 
\begin{align}
(0, 0, \pm {\rm i}|\vec{k}|),\label{eq:CAF_ADM}
\end{align}
where the real-part of all of the CAFs are zero.
Therefore the introduction of the $C^2$-adjusted terms to the evolution equations
changes the constraint propagation equations to a self-decay system. 

More precisely, CAFs depend to $|k|^2$ if $\kappa\neq 0$. 
This indicates that adjusted terms affect to reduce 
high frequency error-growing modes. 
Since we intend not to change the original evolution equations drastically
by adding adjusted terms, we consider only small $\kappa$. 
This limits the robustness of the system to the low frequency error-growing 
modes. Therefore the system may stop due to the low frequency modes, but
the longer evolutions are expected to be obtained. 

\subsection{Detweiler's ADM formulation}
\label{Detweiler_formulation}
We review Detweiler's ADM formulation \cite{Detweiler87} for a comparison
with the $C^2$-adjusted ADM formulation and the standard ADM formulation.
Detweiler proposed an evolution system in order to ensure the decay of the
norm of constraints, $\partial_t C^2<0$.
His system can be treated as one of the adjusted ADM systems and the set of
evolution equations can be written as
\begin{align}
\partial_t\gamma_{i j}
&=\eqref{eq:gamma_standardADMEvolutionEquations}
 +LD_{\gamma i j},
 \label{eq:gammaEvolutionEqations_DetweilerSystem}\\
\partial_t K_{i j}
&=\eqref{eq:extrinsicCurvature_standardADMEvolutionEquations}
+LD_{K i j},
\label{eq:extrinsicCurvatureEvolutionEqations_DetweilerSystem}\\
{\rm where\qquad}
D_{\gamma i j}&\equiv -\alpha^3\gamma_{i j}\mathcal{H},
\label{eq:adjustedGammaDetweilerSystem}\\
D_{K i j}&\equiv
\alpha^3(K_{i j}-(1/3)K\gamma_{i j})\mathcal{H}\nonumber\\
&\quad
+\alpha^2[3(\partial_{(i}\alpha)\delta^k{}_{j)}-(\partial_\ell\alpha)\gamma_{i j}\gamma^{k\ell}]\mathcal{M}_k
\nonumber\\
&\quad
+\alpha^3[\delta^k{}_{(i}\delta^\ell{}_{j)}-(1/3)\gamma_{i j}\gamma^{k\ell}]
D_k\mathcal{M}_\ell,
\label{eq:adjustedKDetweilerSystem}
\end{align}
where $L$ is a constant.
He found that with this particular combination of adjustments, the evolution of
the norm constraints, $C^2$, can be negative definite when we apply the maximal
slicing condition, $K=0$, for fixing the lapse function, $\alpha$.
Note that the effectiveness with other gauge conditions is remain unknown.
The numerical demonstrations with Detweiler's ADM formulation are presented in
\cite{YS01prd, Shinkai09}, and there we can see the drastic improvements for
stability.

The CAFs of Detweiler's ADM formulation on flat background metric are derived as
\cite{YS01prd},
\begin{align}
&\biggl(-(L/2)|\vec{k}|^2,
-(L/2)|\vec{k}|^2,\nonumber\\
&\quad
-(4L/3)|\vec{k}|^2
\pm \sqrt{|\vec{k}|^2\{-1+(4/9)L^2|\vec{k}|^2\}}
\biggr),
\label{eq:CAFs_Detweiler}
\end{align}
which indicates the constraints will damp down if $L>0$, apparently better
feature than the standard ADM formulation.

\section{Numerical Examples}
\label{NumericalExamples}

We demonstrate the damping of constraint violations in numerical evolutions
using the polarized Gowdy-wave spacetime, which is one of the standard tests
for comparisons of formulations in numerical relativity as is known as the
Apples-with-Apples testbeds\cite{Alcubierre_etc04}.
The tests have been used by several groups and were reported in the same
manner (e.g. \cite{Zumbusch09, BB10, KS08}).

The testbeds provide three tests of the solutions of the Einstein equations:
gauge-wave, linear-wave, and Gowdy-wave tests.
Among these tests, we report only on the Gowdy-wave test.
This is because the other two are based on the flat backgrounds and the
violations of constraints are already small, so that the differences of
evolutions between the ADM, $C^2$-adjusted ADM, and Detweiler-ADM are
indistinguishable.

\subsection{Gowdy-wave Testbed}
\label{GOWDY_TEST}
The metric of the polarized Gowdy-wave is given by
\begin{align}
d s^2=t^{-1/2}{\rm e}^{\lambda/2}(-d t^2+d x^2)
+t({\rm e}^P d y^2+{\rm e}^{-P}dz^2),
\end{align}
where $P$ and $\lambda$ are functions of $x$ and $t$.
The time coordinate $t$ is chosen such that time increases as the universe
expands, this metric is singular at $t=0$ which corresponds to the cosmological
singularity.

For simple forms of the solutions, $P$ and $\lambda$, are given by
\begin{align}
P&=J_0(2\pi t)\cos(2\pi x),\\
\lambda&=
-2\pi t J_0(2\pi t)J_1(2\pi t)\cos^2(2\pi x)
+2\pi^2t^2
[J_0^2(2\pi t)
\nonumber\\
&\quad
+J_1^2(2\pi t)]
-(1/2)\{
(2\pi)^2[J_0^2(2\pi)+J_1^2(2\pi)]
\nonumber\\
&\quad
-2\pi J_0(2\pi)J_1(2\pi)
\},
\end{align}
where $J_n$ is the Bessel function.

Following \cite{Alcubierre_etc04}, the new time coordinate $\tau$, which
satisfies the harmonic slicing, is obtained by coordinate transformation as
\begin{align}
t(\tau) = k {\rm e}^{c\tau},
\end{align}
where $k$ and $c$ are arbitrary constants.
We also follow \cite{Alcubierre_etc04} for choosing these constants $k$, $c$
and initial time $t_0 $ as
\begin{align}
k &\sim 9.67076981276405,\quad
c \sim 0.002119511921460,\\
t_0 &=9.87532058290982
\end{align}
in such a way that the lapse function in new time coordinate is unity and
$t=\tau$ at initial time.

We also use following parameters specified in \cite{Alcubierre_etc04},
\begin{itemize}
  \item Simulation domain: $x\in[-0.5,0.5], y=z=0$.
  \item Grid: $x_n = -0.5+(n-(1/2))dx$, $n=1,\cdots,100$,
  where $d x=1/100$.
  \item Time step: $d t=0.25d x$.
  \item Boundary conditions: Periodic boundary condition in
  $x$-direction and planar symmetry in $y$- and $z$-directions.
  \item Gauge conditions: the harmonic slicing and
  $\beta^i=0$.
  \item Scheme: second order iterative Crank-Nicholson.
\end{itemize}
Our code passed convergence tests with the second-order accuracy.

\subsection{Constraint violations and the damping of the violations}
\label{ConstraintViolation}
\begin{figure}[t]
\includegraphics[keepaspectratio=true,width=85mm]{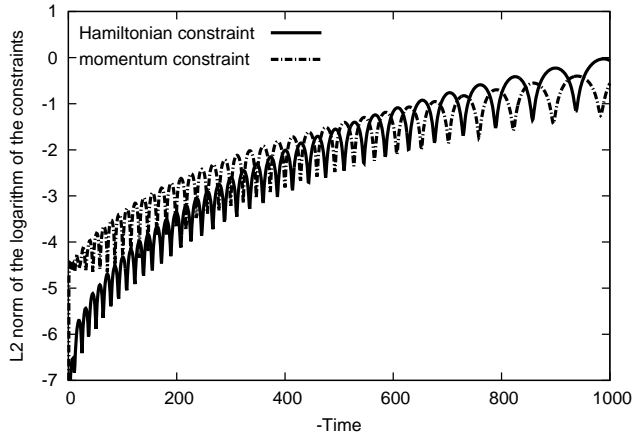}
\caption{
The L2 norm of the Hamiltonian and momentum constraints of the Gowdy-wave
evolution using the standard ADM formulation.
We see that the violation of the momentum constraints is larger initially, and
both violations are growing with time.
\label{fig:standardADM_H_M}}
\end{figure}

\begin{figure}[t]
\includegraphics[keepaspectratio=true,width=85mm]{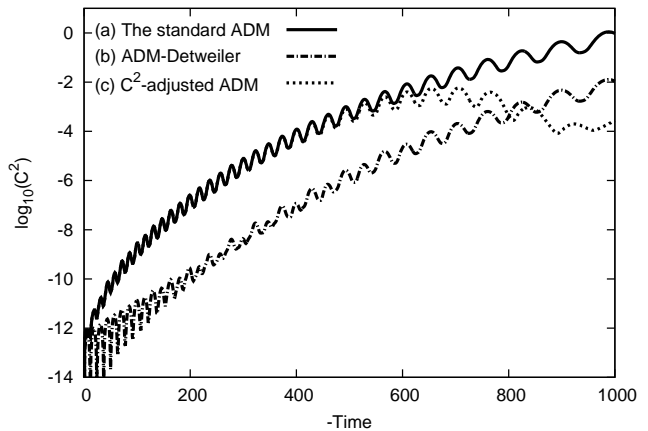}
\caption{
The L2 norm of the constraints, $C^2$, of the polarized Gowdy-wave tests with
ADM and two types of adjusted formulations.
The vertical axis is the logarithm of the $C^2$ and the horizontal axis is
backward time.
The solid line (a) is of the standard ADM formulation.
The dot-dashed line (b)  is the evolution with Detweiler's ADM with
$L=-10^{+1.9}$.
The dotted line (c) is the $C^2$-adjusted ADM with $\kappa_{\gamma}=-10^{-9.0}$
and $\kappa_{K}=-10^{-3.5}$.
We see the lines (a) and (c) almost overlap until $t=-500$, then the case (c)
keeps the L2 norm at the level $\leq 10^{-3}$, while the lines of (a) and (b)
monotonically grow larger with oscillations.
We confirmed this behavior up to $t\simeq -1700$.
\label{fig:ADMformulations}}
\end{figure}

\begin{figure}[t]
\includegraphics[keepaspectratio=true,width=85mm]{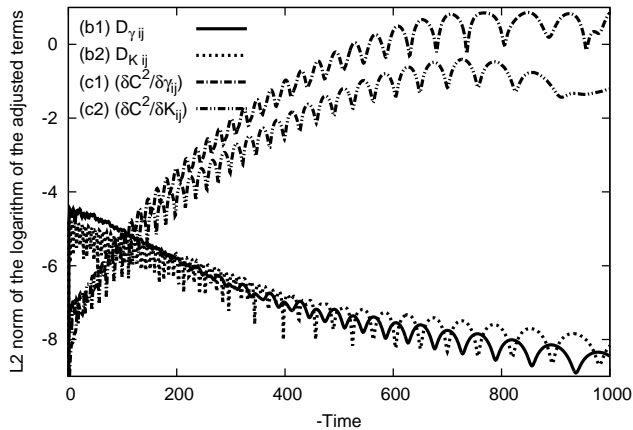}
\caption{
The magnitudes of the adjusted terms in each equations for the evolutions shown
in Figure \ref{fig:ADMformulations}.
The vertical axis is the logarithm of the adjusted terms.
The horizontal axis is backward time.
The lines (b1) and (b2) are the adjusted terms
\eqref{eq:adjustedGammaDetweilerSystem} and \eqref{eq:adjustedKDetweilerSystem}
respectively.
The lines (c1) and (c2) are the adjusted terms
\eqref{eq:appendix_ADM_deltaGamma1} and \eqref{eq:appendix_ADM_deltaK1}
respectively.
We see the adjustments in Detweiler-ADM [the lines (b1) and (b2)] decrease with
time, which indicates that these contributions become less effective.
\label{fig:adjustedTerms}}
\end{figure}

Figure \ref{fig:standardADM_H_M} shows the L2 norm of the Hamiltonian
constraint and momentum constraints with a function of backward time $(-t)$ in
the case of the standard ADM formulation,
\eqref{eq:gamma_standardADMEvolutionEquations}-\eqref{eq:extrinsicCurvature_standardADMEvolutionEquations}.
We see the violations of the momentum constraints are larger than that of the
Hamiltonian constraint at the initial stage, and both grow larger with time.
The behavior is well-known, and the starting point of the formulation problem.

We, then, compare the evolutions with three formulations: (a)
the standard ADM formulation
\eqref{eq:gamma_standardADMEvolutionEquations}-\eqref{eq:extrinsicCurvature_standardADMEvolutionEquations},
(b) Detweiler's formulation
\eqref{eq:gammaEvolutionEqations_DetweilerSystem}-\eqref{eq:extrinsicCurvatureEvolutionEqations_DetweilerSystem},
and (c) the $C^2$-adjusted ADM formulation
\eqref{eq:gamma_c2AdjustedSystemADM}-\eqref{eq:extrinsicCurvature_c2AdjustedSystemADM}.
We tuned the parameters $L$ in (a), and $\kappa_{\gamma i j m n}$ and $\kappa_{K
i j m n}$ in (c) within the expected ranges from the eigenvalue analyses.
In the formulation (c), we set $\kappa_{\gamma i j m n} =
\kappa_{\gamma}\delta_{i m}\delta_{j n}$ and $\kappa_{K i j m n} =
\kappa_{K}\delta_{i m}\delta_{j n}$  for simplicity, and optimized
$\kappa_\gamma$ and $\kappa_K$ in their positive ranges.
We use $L=-10^{+1.9}$ and $(\kappa_\gamma, \kappa_k) = (-10^{-9.0}, -10^{-3.5})$
for the plots, since the violation of constraints are minimized at $t=-1000$ for
those evolutions.
Note that the signatures of $(\kappa_\gamma, \kappa_K)$ and $L$ are reversed
from the expected one in Sec. \ref{GeneralIdea} and Sec.
\ref{ApplicationEinsteinEq}, respectively, since we integrate time backward.

We plot the L2 norms of $C^2$ of these three formulations in Figure
\ref{fig:ADMformulations}.
We see the constraint violations of  (a)(the standard ADM formulation) and
(b)(Detweiler's formulation) grow larger with time, while that of
(c)($C^2$-adjusted ADM formulation) almost coincide with (a) until $t=-500$,
then the violation of (c) begins smaller than (a).
The L2 violation level of (c), then, keeps its magnitude at most $O(10^{-3})$,
while those of (a) and (b) monotonically grow larger with oscillations.
Figure \ref{fig:ADMformulations} shows up to $t=-1000$, but we confirmed this
behavior up to $t=-1700$.

Figure \ref{fig:ADMformulations} tells us that the effects of Detweiler's
adjustment appear at the initial stage, while $C^2$-adjustment contributes at
the later stage.
The time difference can be seen also from the magnitudes of adjustment terms in
each evolution equations, which we show in Figure \ref{fig:adjustedTerms}.
The lines (b1), (b2), (c1), and (c2) are the norms of $D_{\gamma ij}$ in
\eqref{eq:adjustedGammaDetweilerSystem}, $D_{K ij}$ in
\eqref{eq:adjustedKDetweilerSystem},
$\delta C^2/\delta \gamma_{i j}$ in \eqref{eq:appendix_ADM_deltaGamma1},
and $\delta C^2/\delta K_{i j}$ in \eqref{eq:appendix_ADM_deltaK1},
respectively.

We see that the L2 norms of the adjusted terms of Detweiler's ADM
formulation, $D_{\gamma i j}$ and $D_{K i j}$, decrease, while that of
the $C^2$-adjusted ADM formulation increase.
If the magnitudes of the adjusted terms are smaller, the effects of the
constraint damping become small.
Therefore, the L2 norm of $C^2$ of Detweiler's ADM formulation are
not damped down in the later stage in Figure \ref{fig:ADMformulations}.

One possible explanation for the weak effect of Detweiler's adjustment in the
later stage is the existence of the lapse function, $\alpha$ (and $\alpha^2$,
$\alpha^3$), in the adjusted terms in
\eqref{eq:adjustedGammaDetweilerSystem}-\eqref{eq:adjustedKDetweilerSystem}.
The Gowdy-wave testbed is the evolution to the initial singularity of the
space-time, and the lapse function becomes smaller with evolution.
Note that in previous works \cite{YS01prd, Shinkai09}, we see that the
constraint violations are damped down in the simulation with Detweiler's ADM
formulation, where the lapse function, $\alpha$, is adopted by the geodesic condition.

\begin{figure}[t]
\includegraphics[keepaspectratio=true,width=85mm]{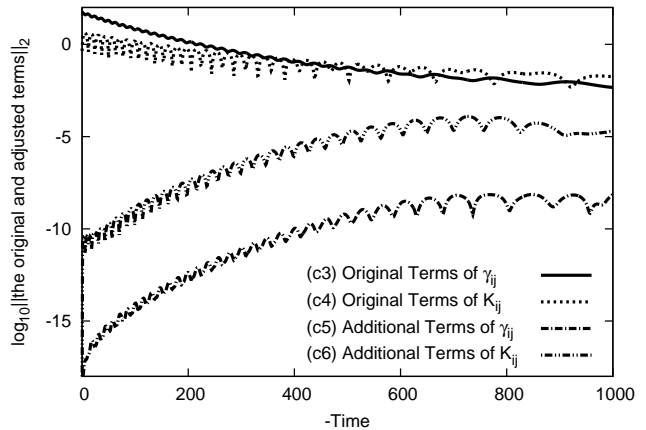}
\caption{
Comparison of the magnitude of the original terms and the adjusted terms
of the $C^2$-adjusted ADM formulation, \eqref{eq:gamma_c2AdjustedSystemADM}-\eqref{eq:extrinsicCurvature_c2AdjustedSystemADM}. 
The lines (c3) and (c4) are the L2 norm of the original terms [the evolution equations of 
$g_{i j}$ and $K_{i j}$, \eqref{eq:gamma_standardADMEvolutionEquations} and \eqref{eq:extrinsicCurvature_standardADMEvolutionEquations}], respectively. 
The lines (c5) and (c6) are the L2 norm of the adjusted terms, 
which is the second terms of the right-hand side of \eqref{eq:gamma_c2AdjustedSystemADM} and \eqref{eq:extrinsicCurvature_c2AdjustedSystemADM}, respectively. 
We see the adjusted terms are ``tiny", compared with the original terms. 
 \label{fig:DifferenceOriginalAndAdjusted}
}
\end{figure}

In Figure \ref{fig:DifferenceOriginalAndAdjusted}, we plotted the magnitude of the original terms and 
the adjusted terms of $C^2$-adjusted ADM formulation; 
the first and second terms in \eqref{eq:gamma_c2AdjustedSystemADM}
and \eqref{eq:extrinsicCurvature_c2AdjustedSystemADM}.  We find that
there is $O(10^2)$--$O(10^5)$ of differences between them. 
Therefore, we conclude that the adjustments do not disturb the original
ADM formulation, but control the violation of the constraints.
We may understand that higher derivative terms in 
\eqref{eq:appendix_ADM_deltaGamma1} and \eqref{eq:appendix_ADM_deltaK1} 
work as artificial viscosity terms in numerics.
\subsection{Parameter dependence of the $C^2$-adjusted ADM formulation}
\label{parameter_dependence}

\begin{figure*}[t]
\includegraphics[keepaspectratio=true,width=170mm]{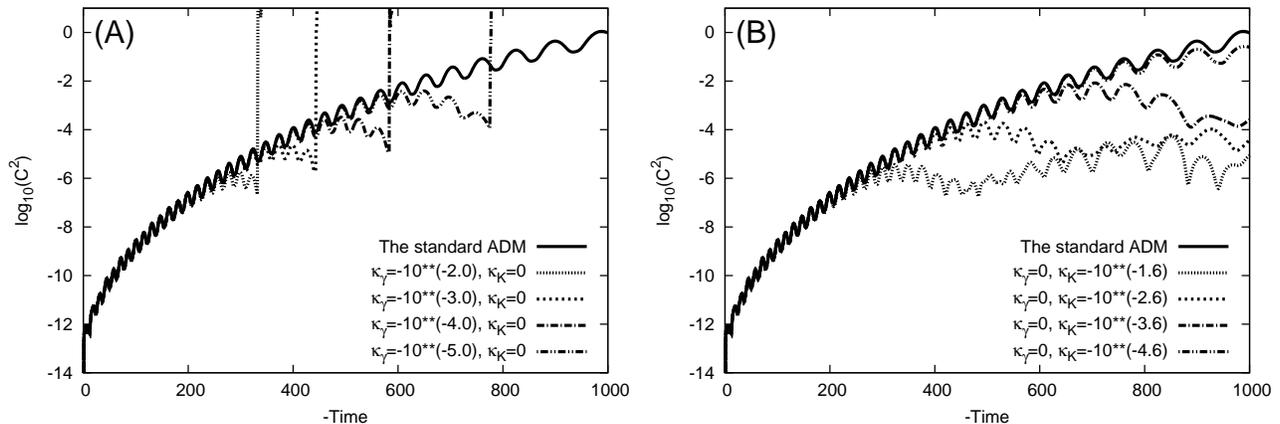}
\caption{
Parameter dependence of the $C^2$-adjusted ADM formulation.
The vertical axis is the logarithm of the $C^2$ and the horizontal axis is
backward time.
The left panel (A) is the evolutions with $\kappa_K=0$ and
$\kappa_\gamma=-10^{-2.0}, -10^{-3.0}, -10^{-4.0}, -10^{-5.0}$.
The right panel (B) is the cases with $\kappa_\gamma=0$ and
$\kappa_K=-10^{-1.6}, -10^{-2.6}, -10^{-3.6},-10^{-4.6}$.
In (A), we see that the simulations stop soon after the constraint dumping
effect appears.
In (B), we see that the simulations continue with constraint-damping effects.
\label{fig:ParameterDependence}}
\end{figure*}

There are two parameters, $\kappa_\gamma$ and $\kappa_K$, in the $C^2$-adjusted
ADM formulation and we next study the sensitivity of these two on the damping
effect to the constraint violation.

Figure \ref{fig:ParameterDependence} shows the dependences on $\kappa_\gamma$
and $\kappa_K$.
In Figure \ref{fig:ParameterDependence} (A), we fix $\kappa_K=0$ and change
$\kappa_\gamma$.
In Figure \ref{fig:ParameterDependence} (B), we fix $\kappa_\gamma=0$ and change
$\kappa_K$.
In Figure \ref{fig:ParameterDependence} (A), we see that all the
simulations stop soon after the damping effect appears.
On the other hand, in Figure
\ref{fig:ParameterDependence} (B), we see that the simulations continue with
constraint-damping effects.
These results suggest $\kappa_K\neq 0$ or $\kappa_\gamma = 0$ is essential to
keep the constraint-damping effects.

We think the trigger for stopping evolutions in the cases of Figure
\ref{fig:ParameterDependence} (A) (when $\kappa_K=0$) is the term
$\mathbb{H}_5{}^{a b c d}(\partial_a\partial_b\partial_c\partial_d\mathcal{H})$
which appears in the constraint propagation equation of the Hamiltonian
constraint, \eqref{eq:app_CPHC2}.
We evaluated and checked each terms and found that $\mathbb{H}_5{}^{a b c d}$
exponentially grows in time and dominates the other terms in
\eqref{eq:app_CPHC2} before the simulation stops.
Since $\mathbb{H}_5{}^{a b c d}$ is consists of $\gamma^{i j}\gamma^{m n}$ [see
\eqref{eq:appendix_ADM_H3} and \eqref{eq:h5_hamiltonianCP_withC2}], the time
backward integration of Gowdy spacetime makes this term disastrous.
So that, in this Gowdy testbed, the cases $\kappa_\gamma=0$ reduce this trouble
and keep the evolution with constraint-damping effects.

The sudden stops of evolutions in Figure \ref{fig:ParameterDependence} (A) can
be interpreted due to a non-linear growth of ``constraint shocks", since the adjusted terms are highly non-linear
\footnote{We appreciate the anonymous referee for pointing out this issue.}.
The robustness against a constraint-shock is hard to be proved,
but the continuous evolution cases in Figure \ref{fig:ParameterDependence} (B)
may show that a remedial example is available by tuning parameters.

\section{Summary}
\label{Summary}
In order to construct a robust and stable formulation,
we proposed a new set of evolution equations, which we call the $C^2$-adjusted
ADM formulation.
We applied the adjusting method suggested by Fiske \cite{Fiske04} to the ADM
formulation.
We obtained the evolution equations as
\eqref{eq:appendix_ADM_deltaGamma1}-\eqref{eq:appendix_ADM_deltaK1}
and the constraint propagation equations, \eqref{eq:app_CPHC2} and
\eqref{eq:app_CPMC2}, and also discussed the constraint propagation of this
system . We analyzed the constraint amplification factors (CAFs) on the flat background,
and confirmed that all of the CAFs have negative real-part
which indicate the damping of the constraint violations.
We, then,  performed numerical tests with the polarized
Gowdy-wave and showed the damping of the constraint violations as expected.

There are two advantages of the $C^2$-adjusted system.
One is that we can uniquely determine the form of the adjustments.
The other is that we can specify the effective signature of the coefficiencies
(Lagrange multipliers) independent on the background. (The term ``effective"
means that the system has the property of the damping constraint violations).
In the previous our study \cite{YS01prd}, we systematically examined several
combinations of adjustments to the ADM evolution equations, and discuss the
effective signature of those Lagrange multipliers using CAFs as the guiding
principle.
However, the $C^2$-adjusted idea, \eqref{eq:adjutedGeneralEvolveEquations},
automatically includes this guiding principle.
We confirm this fact using CAF-analysis on the flat background.

The $C^2$-adjusted idea is the one of the useful ideas to decide the adjustments
with theoretical logic.
We are now applying this idea also to the BSSN formulation which will be
presented elsewhere near future.

We performed the simulation with the $C^2$-adjusted ADM formulation on
the Gowdy-wave spacetime and confirmed the effect of the constraint dumping.
We investigated the parameter dependencies and found that the
constraint-damping effect does not continue due to one of the adjusted terms.
We also found that the Detweiler's adjustment \cite{Detweiler87} is not so
effective against constraint violations on this spacetime.
Up to this moment, we do not yet know how to choose the ranges of parameters
which are suitable to damp the constraint violations unless the simulations are actually performed.

It would be helpful if there are methods to monitor the order of constraint
violations and to maintain them by tuning the Lagrange multipliers
automatically.
Such an implementation would make numerical relativity more friendly to the
beginners.
Applications of the controlling theories in this direction are in progress.

\begin{acknowledgments}
This work was supported partially by the Grant-in-Aid for
Scientific Research Fund of Japan Society of the Promotion of Science,
No. 22540293 (HS).
Numerical computations were carried out on Altix3700 BX2 at YITP in Kyoto University,
and on the RIKEN Integrated Cluster of Clusters (RICC).
\end{acknowledgments}

\appendix

\section{The additional $C^2$-adjusted terms}
\label{Appendix_adjustTermsC2}
The adjusted terms $\delta C^2/\delta \gamma_{m n}$ and $\delta
C^2/\delta K_{m n}$ in
\eqref{eq:gamma_c2AdjustedSystemADM}-\eqref{eq:extrinsicCurvature_c2AdjustedSystemADM}
are written as
\begin{widetext}
\begin{align}
\frac{\delta C^2}{\delta \gamma_{m n}}
&=
2H_1{}^{m n}\mathcal{H}
-2(\partial_\ell H_2{}^{m n \ell})\mathcal{H}
-2 H_2{}^{m n \ell}(\partial_\ell \mathcal{H})
+2(\partial_k\partial_\ell H_3{}^{m n k \ell})\mathcal{H}
+4(\partial_\ell H_3{}^{m n k \ell})(\partial_k\mathcal{H})
+2 H_3{}^{m n k \ell}(\partial_k\partial_\ell \mathcal{H})
\nonumber\\&\quad
+2 M_{1 i}{}^{m n}\mathcal{M}^i
-2(\partial_\ell M_{2 i}{}^{m n \ell})\mathcal{M}^i
-2 M_{2 i}{}^{m n \ell} (\partial_\ell \mathcal{M}^i)
-\mathcal{M}^m\mathcal{M}^n,
\label{eq:appendix_ADM_deltaGamma1}
\\
\frac{\delta C^2}{\delta K_{m n}}
&=
2H_4{}^{m n}\mathcal{H}
+2M_{3 i}{}^{m n}\mathcal{M}^i
-2(\partial_\ell M_{4 i}{}^{m n \ell})\mathcal{M}^i
-2 M_{4 i}{}^{m n \ell}(\partial_\ell \mathcal{M}^i),
\label{eq:appendix_ADM_deltaK1}
\end{align}
\end{widetext}
where
\begin{align}
H_1^{m n}&=
-2R^{m n}
+{}^{(3)}\Gamma^m{}^{(3)}\Gamma^n
-{}^{(3)}\Gamma^{m e b}{}^{(3)}\Gamma^n{}_{e b}
\nonumber\\&\quad
-2KK^{m n}
+2K^m{}_jK^{n j},
\label{eq:appendix_ADM_H1}
\\
H_2^{i m n}
&=
-\gamma^{\ell m}{}^{(3)}\Gamma^n
-\gamma^{\ell n}{}^{(3)}\Gamma^m
+\gamma^{m n}{}^{(3)}\Gamma^\ell
+{}^{(3)}\Gamma^{n m \ell}
\nonumber\\&\quad
+{}^{(3)}\Gamma^{m n \ell}
-{}^{(3)}\Gamma^{\ell n m},
\label{eq:appendix_ADM_H2}
\\
H_3^{i j m n}
&=
\frac{1}{2}\gamma^{m \ell}\gamma^{n k}
+\frac{1}{2}\gamma^{k m}\gamma^{n \ell}
-\gamma^{k \ell}\gamma^{m n}
,
\label{eq:appendix_ADM_H3}
\\
H_4^{m n}
&=
2\gamma^{m n}K-2K^{m n},
\label{eq:appendix_ADM_H4}
\\
M_1^{m n}{}_i
&=
-\frac{1}{2}K_{\ell i,j}\gamma^{j m}\gamma^{\ell n}
-\frac{1}{2}K_{\ell i,j}\gamma^{j n}\gamma^{\ell m}
+\frac{1}{2}{}^{(3)}\Gamma^nK^m{}_i
\nonumber\\&\quad
+\frac{1}{2}{}^{(3)}\Gamma^mK^n{}_i
+{}^{(3)}\Gamma^{a m n} K_{a i}
-\frac{1}{2}K^{m c}\gamma^{n b}\gamma_{b c,i}
\nonumber\\&\quad
-\frac{1}{2}K^{n c}\gamma^{m b}\gamma_{b c,i}
+K_{a b,i}\gamma^{a m}\gamma^{b n},
\label{eq:appendix_ADM_M1}
 \end{align}
 \begin{align}
M_2^{\ell m n}{}_i
&=
-\frac{1}{2}\gamma^{n\ell}K^m{}_i
-\frac{1}{2}\gamma^{m\ell}K^n{}_i
+\frac{1}{2}\gamma^{m n}K^\ell{}_i
\nonumber\\&\quad
+\frac{1}{2}K^{n m}\delta^\ell{}_i,
\label{eq:appendix_ADM_M2}
\\
M_3^{m n}{}_i
&=
-\frac{1}{2}{}^{(3)}\Gamma^m\delta^n{}_i
-\frac{1}{2}{}^{(3)}\Gamma^n\delta^m{}_i
+\frac{1}{2}\gamma^{n a}\gamma^{m b}\gamma_{a b,i},
\label{eq:appendix_ADM_M3}
\\
M_4^{\ell m n}{}_i
&=
\frac{1}{2}\gamma^{\ell m}\delta^n{}_i
+\frac{1}{2}\gamma^{\ell n}\delta^m{}_i
-\gamma^{m n}\delta^\ell{}_i,
\label{eq:appendix_ADM_M4}
\end{align}
$H_1^{m n}$, $H_2^{i m n}$, $H_3^{i j m n}$, $H_4^{m n}$, $M_1^{m n}{}_i$,
$M_2^{j m n}{}_i$, $M_3^{m n}{}_i$, $M_4^{j m n}{}_i$ are the same with the
appendix of \cite{SY02} if $(m,n)=(n,m)$.
\begin{widetext}
\section{The constraint propagation equations of $C^2$-adjusted
 ADM formulation}
\label{Appendix_constriantPropagation_C2ADM}
The propagation equation of the Hamiltonian constraint with $C^2$-adjusted ADM
formulation can be written as
\begin{align}
\partial_t\mathcal{H}
&=
\mathbb{H}_1\mathcal{H}
+\mathbb{H}_{2}{}^a(\partial_a \mathcal{H})
+\mathbb{H}_{3}{}^{a b}(\partial_a\partial_b \mathcal{H})
+\mathbb{H}_{4}{}^{a b c}(\partial_a\partial_b \partial_c\mathcal{H})
+\mathbb{H}_{5}{}^{a b c d}(\partial_a\partial_b
\partial_c\partial_d\mathcal{H})
+\mathbb{H}_{6 a} \mathcal{M}^a
+\mathbb{H}_{7 a}{}^b(\partial_b \mathcal{M}^{a})
\nonumber\\&\quad
+\mathbb{H}_{8 a}{}^{b c}(\partial_b \partial_c \mathcal{M}^{a})
+\mathbb{H}_{9 a}{}^{b c d}(\partial_b \partial_c\partial_d
 \mathcal{M}^{a}),
\label{eq:app_CPHC2}
\end{align}
where
\begin{align}
\mathbb{H}_1
&=
2\alpha K
 -2\kappa_{\gamma m n i j}\biggl\{
 H_1{}^{m n}H_1^{i j}
 -H_1{}^{m n}(\partial_c H_2{}^{i j c})
 +H_1{}^{m n}(\partial_d\partial_c H_3{}^{i j d c})
 +H_2{}^{m n\ell}(\partial_\ell H_1{}^{i j})
 -H_2{}^{m n\ell}(\partial_\ell \partial_c
 H_2{}^{i j c})
  \nonumber\\&\qquad
 +H_2{}^{m n\ell}(\partial_\ell \partial_d\partial_c H_3^{i j d c})
 +H_3{}^{m n k\ell}(\partial_k\partial_\ell H_1{}^{i  j})
 -H_3{}^{m n k\ell}(\partial_k\partial_\ell \partial_c H_2{}^{i j
 c})
 +H_3{}^{m n k\ell}(\partial_k\partial_\ell \partial_d \partial_c H_3{}^{i j d
 c})\biggr\}
 \nonumber\\&\quad
 -2(\partial_\ell\kappa_{\gamma m n i j})\biggl\{
 H_2{}^{m n \ell}H_1^{i j}
 -H_2{}^{m n \ell}(\partial_c H_2{}^{i j c})
 +H_2{}^{m n \ell}(\partial_d\partial_c H_3{}^{i j d c})
 +2H_3{}^{m n \ell k}(\partial_k H_1{}^{i j})
 \nonumber\\&\qquad
 -2H_3{}^{m n \ell k}(\partial_k \partial_c H_2{}^{i j c})
 +2H_3{}^{m n \ell k}(\partial_k \partial_d\partial_c H_3^{i j d c})
 \biggr\}
 -2(\partial_k\partial_\ell\kappa_{\gamma m n i j})
 \biggl\{
 H_3{}^{m n k \ell}H_1^{i j}
 -H_3{}^{m n k \ell}(\partial_c H_2{}^{i j c})
\nonumber\\&\qquad
 +H_3{}^{m n k \ell}(\partial_d\partial_c H_3{}^{i j d c})
 \biggr\}
 -2\kappa_{K m n i j }H_4{}^{m n}H_4{}^{i j},
 \label{eq:h1_hamiltonianCP_withC2}
\\
\mathbb{H}_2{}^a
&=
\beta^a
 -2\kappa_{\gamma m n i j}
 \biggl\{
 -H_1{}^{m n}H_2{}^{i j a}
 +2H_1{}^{m n}(\partial_c H_3{}^{i j a c})
 +H_2{}^{m n a}H_1{}^{i j}
-H_2{}^{m n a}(\partial_c H_2{}^{i j c})
-H_2{}^{m n\ell}(\partial_\ell H_2{}^{i j a})
\nonumber\\&\qquad
+H_2{}^{m n a}(\partial_d\partial_c H_3{}^{i j d c})
+2H_2{}^{m n\ell}(\partial_\ell \partial_c H_3{}^{i j a c})
 +H_3{}^{m n a\ell}(\partial_\ell H_1{}^{i j})
 +H_3{}^{m n k a}(\partial_k H_1{}^{i j})
 -H_3{}^{m n a \ell}(\partial_\ell \partial_c H_2{}^{i j c})
\nonumber\\&\qquad
 -H_3{}^{m n k a}(\partial_k\partial_c H_2{}^{i j c})
 -H_3{}^{m n k\ell}(\partial_k\partial_\ell H_2{}^{i j a})
 +H_3{}^{m n a\ell}(\partial_\ell \partial_d \partial_c H_3^{i j d
 c})
 +H_3{}^{m n k a}(\partial_k\partial_d\partial_c H_3{}^{i j d
 c})
\nonumber \\&\qquad
 +2H_3{}^{m n k\ell}(\partial_k \partial_\ell \partial_c H_3{}^{i j a c})
 \biggr\}
 -2(\partial_\ell
 \kappa_{\gamma m n i j})
 \biggl\{
 -H_2{}^{m n \ell}H_2{}^{i j a}
 +2H_2{}^{m n \ell}(\partial_c H_3{}^{i j a c})
 +2H_3{}^{m n \ell a}H_1{}^{i j}
\nonumber\\&\qquad
 -2H_3{}^{m n \ell a}(\partial_c H_2{}^{i j c})
-2H_3{}^{m n \ell k}(\partial_kH_2{}^{i j a})
+2H_3{}^{m n \ell a}(\partial_d\partial_c H_3{}^{i j d c})
+4H_3{}^{m n \ell k}(\partial_k \partial_c H_3{}^{i j a c})\biggr\}\nonumber\\
 &\quad
 -2(\partial_k\partial_\ell\kappa_{\gamma m n i j})
 \biggl\{
 -H_3{}^{m n k
 \ell}H_2{}^{i j a}
 +2H_3{}^{m n k
 \ell}(\partial_c H_3{}^{i j a c})
 \biggr\},
 \label{eq:h2_hamiltonianCP_withC2}
 \\
\mathbb{H}_3{}^{a b}
&=
-2\kappa_{\gamma m n i j}
 \biggl\{
 H_1{}^{m n}H_3{}^{i j a b}
 -H_2{}^{m n a}H_2{}^{i j b}
 +2H_2{}^{m n a}(\partial_c H_3{}^{i j b c})
 +H_2{}^{m n \ell}(\partial_\ell H_3{}^{i j a b})
 +H_3{}^{m n a b}H_1{}^{i j}
\nonumber\\&\qquad
 -H_3{}^{m n a b}(\partial_c H_2{}^{i j c})
 -H_3{}^{m n a \ell}(\partial_\ell H_2{}^{i j b})
 -H_3{}^{m n k a}(\partial_k H_2{}^{i j b})
 +H_3{}^{m n a b}(\partial_d\partial_c H_3{}^{i j d c})
  \nonumber\\&\qquad
 +2H_3{}^{m n a \ell}(\partial_\ell \partial_c H_3{}^{i j b c})
 +2H_3{}^{m n k a}(\partial_k\partial_c H_3{}^{i j b c})
 +H_3{}^{m n k \ell}(\partial_k\partial_\ell H_3{}^{i j a b})
 \biggr\}
 \nonumber\\&\quad
 -2(\partial_\ell \kappa_{\gamma m n i j})\biggl\{
 H_2{}^{m n \ell}H_3{}^{i j  a b}
 -2H_3^{m n \ell a}H_2{}^{i j b}
 +4H_3^{m n \ell a}(\partial_c H_3{}^{i j b c})
 +2H_3^{m n \ell k}(\partial_k H_3{}^{i j a b})\biggr\}
 \nonumber\\&\quad
 -2(\partial_k\partial_\ell \kappa_{\gamma m n i j})
 H_3{}^{m n k \ell}H_3{}^{i j a b},
 \label{eq:h3_hamiltonianCP_withC2}
 \\
\mathbb{H}_4{}^{a b c}
&=
-2\kappa_{\gamma m n i j}\biggl\{
 H_2{}^{m n a}H_3{}^{i j b c}
 -H_3{}^{m n a b}H_2{}^{i j c}
 +2H_3{}^{m n a b}(\partial_e H_3{}^{i j c e})
 +H_3{}^{m n a\ell}(\partial_\ell H_3{}^{i j b c})
 +H_3{}^{m n k a}(\partial_k H_3{}^{i j b c})
 \biggr\}
 \nonumber\\&\quad
 -4(\partial_k \kappa_{\gamma m n i j})H_3{}^{m n k a}H_3{}^{i j b
 c},
 \label{eq:h4_hamiltonianCP_withC2}
 \\
\mathbb{H}_5{}^{a b c d}
&=
-2\kappa_{\gamma m n i j}H_3{}^{m n a b}
H_3{}^{i j c d},
\label{eq:h5_hamiltonianCP_withC2}
\\
\mathbb{H}_{6 a}
&=
 -2\alpha {}^{(3)}\Gamma^b{}_{b a}
-4\alpha_{,a}
-\kappa_{\gamma m n i j}
 \biggl\{
 2H_1{}^{m n}M_{1 a}{}^{i j}
 -2H_1{}^{m n}(\partial_d M_{2 a}{}^{i j d})
 -H_1{}^{m n} \mathcal{M}^{(i}\delta^{j)}{}_a
 +2H_2{}^{m n\ell}(\partial_\ell M_{1 a}{}^{i j})
 \nonumber\\&\qquad
 -2H_2{}^{m n\ell}(\partial_\ell \partial_d M_{2 a}{}^{i j d})
 +2H_3{}^{m n k\ell}(\partial_k \partial_\ell M_{1 a}{}^{i j})
 -2H_3{}^{m n k\ell}(\partial_k \partial_\ell \partial_d M_{2 a}{}^{i j d})
 \biggr\}
 \nonumber\\&\quad
 -(\partial_\ell
 \kappa_{\gamma m n i j})
 \biggl\{
 2H_2{}^{m n \ell}M_{1 a}{}^{i j}
 -2H_2{}^{m n \ell}(\partial_d M_{2 a}{}^{i j d})
 -H_2{}^{m n \ell} \mathcal{M}^{(i}\delta^{j)}{}_a
 +4H_3{}^{m n \ell k}(\partial_k M_{1 a}{}^{i j})
 \nonumber\\&\qquad
 -4H_3{}^{m n \ell k}(\partial_k \partial_d M_{2 a}{}^{i j d})
\biggr\}
-(\partial_k\partial_\ell\kappa_{\gamma m n i j})
 \left\{
 2H_3{}^{m n k\ell}M_{1 a}{}^{i j}
 -2H_3{}^{m n k\ell}(\partial_d M_{2 a}{}^{i j d})
 -H_3{}^{m n k\ell} \mathcal{M}^{(i}\delta^{j)}{}_a
 \right\}
 \nonumber\\&\quad
 -\kappa_{K m n i j}\biggl\{
 2H_4{}^{m n}M_{3 a}{}^{i j}
 -2H_4{}^{m n}(\partial_\ell M_{4 a}{}^{i j \ell})
 \biggr\},
\label{eq:H6_hamiltonianCP_withC2}
\end{align}
\begin{align}
\mathbb{H}_{7 a}{}^b
&=
-2\alpha \delta^b{}_a
-\kappa_{\gamma m n i j}
 \biggl\{
 -2H_1{}^{m n}M_{2 a}{}^{i j b}
 +2H_2{}^{m n b}M_{1 a}{}^{i j}
 -2H_2{}^{m n b}(\partial_d M_{2 a}{}^{i j d})
 -2H_2{}^{m n\ell}(\partial_\ell M_{2 a}{}^{i j b})
 \nonumber\\&\qquad
 -H_2{}^{m n b} \mathcal{M}^j\delta^i{}_a
 -H_2{}^{m n b} \mathcal{M}^i\delta^j{}_a
 +2H_3{}^{m n b\ell}(\partial_\ell M_{1 a}{}^{i j})
 +2H_3{}^{m n k b}(\partial_k M_{1 a}{}^{i j})
 -2H_3{}^{m n b\ell}(\partial_\ell\partial_d M_{2 a}{}^{i j
 d})
  \nonumber\\&\qquad
 -2H_3{}^{m n k b}(\partial_k \partial_d M_{2 a}{}^{i j
 d})
 -2H_3{}^{m n k\ell}(\partial_k\partial_\ell M_{2 a}{}^{i j
 b})
 -H_3{}^{m n b\ell}(\partial_\ell \mathcal{M}^{(i})\delta^{j)}{}_a
 -H_3{}^{m n b\ell}(\partial_\ell \mathcal{M}^{(j})\delta^{i)}{}_a
 \biggr\}\nonumber\\
 &\quad
 -(\partial_\ell
 \kappa_{\gamma m n i j})
 \biggl\{
 -2H_2{}^{m n \ell}M_{2 a}{}^{i j b}
 +4H_3{}^{m n \ell b}M_{1 a}{}^{i j}
 -4H_3{}^{m n \ell b}(\partial_d M_{2 a}{}^{i j d})
 -4H_3{}^{m n \ell k}(\partial_k M_{2 a}{}^{i j b})
 \nonumber\\&\qquad
 -2H_3{}^{m n \ell b} \mathcal{M}^j\delta^i{}_a
 -2H_3{}^{m n \ell b} \mathcal{M}^i\delta^j{}_a
 \biggr\}
 +2(\partial_k\partial_\ell \kappa_{\gamma m n i j})H_3{}^{m n k\ell}M_{2
 a}{}^{i j b}
 +2\kappa_{K m n i j}H_4{}^{m n}M_{4 a}{}^{i j b},
\label{eq:h7_hamiltonianCP_withC2}
\\
\mathbb{H}_{8a}{}^{b c}
&=
-\kappa_{\gamma m n i j}
 \biggl\{
 -2H_2{}^{m n b}M_{2 a}{}^{i j c}
 +2H_3{}^{m n b c}M_{1 a}{}^{i j}
 -2H_3{}^{m n b c}(\partial_d M_{2 a}{}^{i j d})
 -2H_3{}^{m n b\ell}(\partial_\ell M_{2 a}{}^{i j c})
 \nonumber\\&\qquad
 -2H_3{}^{m n k b}(\partial_k M_{2 a}{}^{i j c})
 -H_3{}^{m n b c} \mathcal{M}^j\delta^i{}_a
 -H_3{}^{m n b c} \mathcal{M}^i\delta^j{}_a
 \biggr\}
 +4(\partial_k \kappa_{\gamma m n i j})H_3{}^{m n k b}
 M_{2 a}{}^{i j c},
 \label{eq:h8_hamiltonianCP_withC2}
\\
\mathbb{H}_{9 a}{}^{b c d}
&=
2\kappa_{\gamma m n i j }H_3{}^{m n b c}M_{2 a}{}^{i j d}.
\label{eq:h9_hamiltonianCP_withC2}
\end{align}
The propagation equation of the momentum constraint with $C^2$-adjusted ADM
formulation  can be written as
\begin{align}
\partial_t  \mathcal{M}_a
&=
\mathbb{M}_{1 a}\mathcal{H}
+\mathbb{M}_{2 a}{}^b(\partial_b \mathcal{H})
+\mathbb{M}_{3 a}{}^{b c}(\partial_b\partial_c \mathcal{H})
+\mathbb{M}_{4 a}{}^{b c d}(\partial_b\partial_c\partial_d \mathcal{H})
+\mathbb{M}_{5 a b}\mathcal{M}^b
+\mathbb{M}_{6 a b}{}^c(\partial_c\mathcal{M}^b)
+\mathbb{M}_{7 a b}{}^{c d}(\partial_c\partial_d\mathcal{M}^b),
 \label{eq:app_CPMC2}
\end{align}
where
\begin{align}
\mathbb{M}_{1 a}
&=-\alpha_{,a}
-2\kappa_{\gamma m n i j}
\biggl\{
M_{1 a}{}^{m n}H_1{}^{i j}
-M_{1 a}{}^{m n}(\partial_c H_2{}^{i j c})
+M_{1 a}{}^{m n}(\partial_d\partial_c H_3{}^{i j d c})
 +M_{2 a}{}^{m n\ell}(\partial_\ell H_1{}^{i j})
 \nonumber\\&\qquad
 -M_{2 a}{}^{m n\ell}(\partial_\ell \partial_c H_2{}^{i j c})
 +M_{2 a}{}^{m n\ell}(\partial_\ell \partial_d \partial_c H_3{}^{i j d c})
 \biggr\}
-2(\partial_\ell \kappa_{\gamma m n i j})
\biggl\{
M_{2 a}{}^{m n \ell}H_1{}^{i j}
-M_{2 a}{}^{m n \ell}(\partial_c H_2{}^{i j c})
\nonumber\\&\qquad
+M_{2 a}{}^{m n \ell}(\partial_d\partial_c H_3{}^{i j d c})
\biggr\}
-2\kappa_{K m n i j}
\biggl\{
M_{3 a}{}^{m n}H_4{}^{i j}
+M_{4 a}{}^{m n \ell}(\partial_\ell H_4{}^{i j})
\biggr\}
-2(\partial_\ell \kappa_{K m n i j})M_{4 a}{}^{m n \ell}H_4{}^{i j},
\label{eq:M1_momentumCP_withC2}
\\
\mathbb{M}_{2 a}{}^b
&=
-\frac{1}{2}\alpha \delta_a{}^b
 -2\kappa_{\gamma m n i j}
\biggl\{
-M_{1 a}{}^{m n}H_2{}^{i j b}
+2M_{1 a}{}^{m n}(\partial_c H_3{}^{i j b c})
+M_{2 a}{}^{m n b}H_1{}^{i j}
-M_{2 a}{}^{m n b}(\partial_c H_2{}^{i j c})
\nonumber\\&\qquad
-M_{2 a}{}^{m n\ell}(\partial_\ell H_2{}^{i j b})
+M_{2 a}{}^{m n b}(\partial_d\partial_c H_3{}^{i j d c})
+2M_{2 a}{}^{m n\ell}(\partial_\ell \partial_c H_3{}^{i j b c})
\biggr\}
\nonumber\\&\quad
-2(\partial_\ell \kappa_{\gamma m n i j})
\biggl\{
-M_{2 a}{}^{m n \ell}H_2{}^{i j b}
 +2M_{2 a}{}^{m n \ell}(\partial_c H_3{}^{i j b c})
 \biggr\}
  -2\kappa_{K m n i j}M_{4 a}{}^{m n b}
  H_4{}^{i j},
\label{eq:M2_momentumCP_withC2}
\\
\mathbb{M}_{3 a}{}^{b c}
&=-2
\kappa_{\gamma m n i j}
\biggl\{
M_{1 a}{}^{m n}H_3{}^{i j b c}
-M_{2 a}{}^{m n b}H_2{}^{i j c}
+2M_{2 a}{}^{m n b}(\partial_dH_3{}^{i j c d})
+M_{2 a}{}^{m n\ell}(\partial_\ell H_3{}^{i j b c})
\biggr\}
\nonumber\\&\quad
-2(\partial_\ell \kappa_{\gamma m n i j})M_{2 a}{}^{m n \ell}H_3{}^{i j b
c},
\label{eq:M3_momentumCP_withC2}
\\
\mathbb{M}_{4 a}{}^{b c d}
&= -2\kappa_{\gamma m n i j}M_{2 a}{}^{m n b}
 H_3{}^{i j c d},
 \label{eq:M4_momentumCP_withC2}
\\
\mathbb{M}_{5 a b}
&=
\gamma_{m b}\beta^m{}_{,a}
+\beta^\ell\gamma_{a b,\ell}
+\alpha K \gamma_{a b}
-\kappa_{\gamma m n i j}
\biggl\{
2M_{1 a}{}^{m n}M_{1 b}{}^{i j}
-2M_{1 a}{}^{m n}(\partial_d M_{2 b}{}^{i j d})
-M_{1 a}{}^{m n} \mathcal{M}^{(j}\delta^{i)}{}_b
\nonumber\\&\qquad
 +2M_{2 a}{}^{m n\ell}(\partial_\ell M_{1 b}{}^{i j})
 -2M_{2 a}{}^{m n\ell}(\partial_\ell \partial_d M_{2 b}{}^{i j d})
 \biggr\}
 -(\partial_\ell \kappa_{\gamma m n i j})
\biggl\{
2M_{2 a}{}^{m n \ell}M_{1 b}{}^{i j}
-2M_{2 a}{}^{m n \ell}(\partial_d M_{2 b}{}^{i j d})
\nonumber\\&\qquad
-M_{2 a}{}^{m n \ell} \mathcal{M}^{(j}\delta^{i)}{}_b
\biggr\}
 -2\kappa_{K m n i j}
 \biggl\{
 M_{3 a}{}^{m n}M_{3 b}{}^{i j}
 -M_{3 a}{}^{m n}(\partial_\ell M_{4 b}{}^{i j \ell})
 +M_{4 a}{}^{m n\ell}(\partial_\ell M_{3 b}{}^{i j})
 \nonumber\\&\qquad
 -M_{4 a}{}^{m n\ell}(\partial_\ell \partial_d M_{4 b}{}^{i j d})
 \biggr\}
 -2(\partial_\ell \kappa_{K m n i j})
 \biggl\{
 M_{4 a}{}^{m n\ell}M_{3 b}{}^{i j}
 -M_{4 a}{}^{m n\ell}(\partial_d M_{4 b}{}^{i j d})
 \biggr\},
 \label{eq:M5_momentumCP_withC2}
\end{align}
\begin{align}
\mathbb{M}_{6 a b}{}^c
&=\beta^c\gamma_{a b}
-\kappa_{\gamma m n i j}
\biggl\{
-2M_{1 a}{}^{m n}M_{2 b}{}^{i j c}
 +2M_{2 a}{}^{m n c}M_{1 b}{}^{i j}
 -2M_{2 a}{}^{m n c}(\partial_d M_{2 b}{}^{i j d})
 -2M_{2 a}{}^{m n\ell}(\partial_\ell M_{2 b}{}^{i j c})
 \nonumber\\&\qquad
 -M_{2 a}{}^{m n c} \mathcal{M}^j\delta^i{}_b
 -M_{2 a}{}^{m n c} \mathcal{M}^i\delta^j{}_b
 \biggr\}
 +2(\partial_\ell \kappa_{\gamma m n i j})M_{2 a}{}^{m n \ell}
M_{2 b}{}^{i j c}
\nonumber\\&\quad
 -2\kappa_{K m n i j}
 \biggl\{
 -M_{3 a}{}^{m n}M_{4 b}{}^{i j c}
 +M_{4 a}{}^{m n c}M_{3 b}{}^{i j}
 -M_{4 a}{}^{m n c}(\partial_d M_{4 b}{}^{i j d})
 -M_{4 a}{}^{m n\ell}(\partial_\ell M_{4 b}{}^{i j c})
 \biggr\}
 \nonumber\\&\quad
 +2(\partial_\ell \kappa_{K m n i j})M_{4 a}{}^{m
 n\ell}M_{4 b}{}^{i j c},
 \label{eq:M6_momentumCP_withC2}
\\
 \mathbb{M}_{7 a b}{}^{c d}
&=
 2\kappa_{\gamma m n i j}M_{2 a}{}^{m n c}
 M_{2 b}{}^{i j d}
 +2\kappa_{K m n i j}M_{4 a}{}^{m n c}
 M_{4 b}{}^{i j d}.
 \label{eq:M7_momentumCP_withC2}
\end{align}
\end{widetext}




\begin{thebibliography}{500}
\bibitem{ADM62}
R. Arnowitt, S. Deser, and C. W. Misner, in {\it Gravitation: An
Introduction to Current Research}, edited by L. Witten (Wiley,
New York, 1962).
\bibitem{York78}
J. W. York, Jr., in {\it Sources of Gravitational Radiation}, edited
by L. Smarr (Cambridge, 1979); L. Smarr and J. W. York, Jr.,
Phys. Rev. D {\bf 17}, 2529 (1978).
\bibitem{SY00}
H. Shinkai and G. Yoneda,
Class. Quant. Grav. {\bf 17}, 4799 (2000).
\bibitem{SY02gr-qc}
H. Shinkai and G. Yoneda,
gr-qc/0209111 (2002).
\bibitem{Shinkai09}
H. Shinkai,
J. Korean Phys. Soc. {\bf 54}, 2513 (2009).
\bibitem{SN95}
M. Shibata and T. Nakamura,
Phys. Rev. D {\bf 52}, 5428 (1995).
\bibitem{BS98}
T. W. Baumgarte and S. L. Shapiro,
Phys. Rev. D {\bf 59}, 024007 (1998).
\bibitem{PretriusCQG05}
F. Pretorius,
Class. Quant. Grav. {\bf 22}, 425 (2005).
\bibitem{Garfinkle02}
D. Garfinkle,
Phys. Rev. D {\bf 65}, 044029 (2002).
\bibitem{KST01}
L. E. Kidder, M. A. Scheel, and S. A. Teukolsky,
Phys. Rev. D {\bf 64}, 064017 (2001).
\bibitem{BCCKM06}
J. G. Baker, J. Centrella, D.-I. Choi, M. Koppitz, and J. van
Meter, Phys. Rev. Lett. {\bf 96}, 111102 (2006).
\bibitem{CLMZ06}
M. Campanelli, C. O. Lousto, P. Marronetti, and Y. Zlochower,
Phys. Rev. Lett. {\bf 96}, 111101 (2006).
\bibitem{Pretorius05}
F. Pretorius,
Phys. Rev. Lett. {\bf 95}, 121101 (2005).
\bibitem{SBCKMP09}
M. A. Scheel, M. Boyle, T. Chu, L. E. Kidder, K. D. Matthews, and H. P.
Pfeiffer,
Phys. Rev. D {\bf 79}, 024003 (2009).
\bibitem{YS01cqg}
G. Yoneda and H. Shinkai,
Class. Quant. Grav. {\bf 18}, 441-462 (2001).
\bibitem{YS02}
G. Yoneda and H. Shinkai,
Phys. Rev. D {\bf 66}, 124003 (2002).
\bibitem{BLPZ03}
C. Bona, T. Ledvinka, C. Palenzuela, and M. \v{Z}\'a\v{c}ek,
Phys. Rev. D {\bf 67}, 104005 (2003).
\bibitem{GCHM05}
C. Gundlach, G. Calabrese, I. Hinder,
and J. M. Mart\'in-Garc\'ia
Class. Quant. Grav. {\bf 22}, 3767 (2005).
\bibitem{PHK08}
V. Paschalidis, J. Hansen, and A. Khokhlov,
Phys. Rev. D {\bf 78}, 064048 (2008).
\bibitem{BBCKMM08}
J. G. Baker, W. D. Boggs, J. Centrella, B. J. Kelly, S. T. McWilliams, and
J. R. van Meter, Phys. Rev. D {\bf  78}, 044046 (2008).
\bibitem{Pretorius06}
F. Pretorius,
Class. Quant. Grav. {\bf 23} (2006) S529.
\bibitem{YS01prd}
G. Yoneda and H. Shinkai,
Phys. Rev. D {\bf 63}, 124019 (2001).
\bibitem{SY02}
H. Shinkai and G. Yoneda,
Class. Quant. Grav. {\bf 19}, 1027 (2002).
\bibitem{Fiske04}
D. R. Fiske,
Phys. Rev. D {\bf 69}, 047501 (2004).
\bibitem{Detweiler87}
S. Detweiler,
Phys. Rev. D {\bf 35}, 1095 (1987).
\bibitem{Alcubierre_etc04}
M. Alcubierre {\it et al.},
Class. Quant. Grav. {\bf 21}, 589 (2004).
\bibitem{KS08}
K. Kiuchi and H. Shinkai,
Phys. Rev. D {\bf 77}, 044010 (2008).
\bibitem{Zumbusch09}
G. Zumbusch,
Class. Quant. Grav. {\bf 26}, 175011 (2009).
\bibitem{BB10}
C. Bona and C. Bona-Casas,
Phys. Rev. D {\bf 82}, 064008 (2010).
\bibitem{BM92}
C. Bona and J. Mass\'o,
Phys. Rev. Lett. {\bf 68}, 1097 (1992).
\bibitem{GM04}
C. Gundlach and J. M. Mart\'in-Garc\'ia,
Phys. Rev. D {\bf 70}, 044032 (2004).
\bibitem{Frittelli97}
S. Frittelli,
Phys. Rev. D {\bf 55}, 5992 (1997).
\bibitem{Hern00}
S. D. Hern,
gr-qc/0004036 (Dissertation).
\end{thebibliography}
\end{document}